\documentclass[prx, twocolumn, superscriptaddress]{revtex4-1}
\usepackage{amsmath, amssymb}
\usepackage[dvipdfmx]{graphicx}
\usepackage{mathtools}
\usepackage{color}
\usepackage[shortlabels]{enumitem}
\begin{document}

\title{Correlated bursts in temporal networks slow down spreading}
\author{Takayuki Hiraoka}
\email{takayuki.hiraoka@apctp.org}
\affiliation{Asia Pacific Center for Theoretical Physics, Pohang 37673, Republic of Korea}
\author{Hang-Hyun Jo}
\email{hang-hyun.jo@apctp.org}
\affiliation{Asia Pacific Center for Theoretical Physics, Pohang 37673, Republic of Korea}
\affiliation{Department of Physics, Pohang University of Science and Technology, Pohang 37673, Republic of Korea}
\affiliation{Department of Computer Science, Aalto University, Espoo FI-00076, Finland}
\date{\today}

\begin{abstract} 
    Spreading dynamics has been considered to take place in temporal networks, where temporal interaction patterns between nodes show non-Poissonian bursty nature. The effects of inhomogeneous interevent times (IETs) on the spreading have been extensively studied in recent years, yet little is known about the effects of correlations between IETs on the spreading. In order to investigate those effects, we study two-step deterministic susceptible-infected (SI) and probabilistic SI dynamics when the interaction patterns are modeled by inhomogeneous and correlated IETs, i.e., correlated bursts. By analyzing the transmission time statistics in a single-link setup and by simulating the spreading in Bethe lattices and random graphs, we conclude that the positive correlation between IETs slows down the spreading. We also argue that the shortest transmission time from one infected node to its susceptible neighbors can successfully explain our numerical results.
\end{abstract}
\maketitle

\section{Introduction}\label{sec:intro}

Characterizing the interaction structure of constituents of complex systems is of utmost importance to understand the dynamical processes in those systems. The interaction structure has been modeled by a network, where nodes and links denote the constituents and their pairwise interactions, respectively~\cite{Albert2002Statistical, Newman2010Networks}. When the interactions are temporal, one can adopt a framework of temporal networks~\cite{Holme2012Temporal}, where links are considered to exist only at the moment of interaction. Events in the temporal interaction patterns are known to be non-Poissonian or bursty~\cite{Barabasi2005Origin, Karsai2012Universal, Karsai2018Bursty}, e.g., as shown in human communication patterns~\cite{Eckmann2004Entropy, Malmgren2009Universality, Cattuto2010Dynamics, Jo2012Circadian, Rybski2012Communication, Jiang2013Calling, Stopczynski2014Measuring, Panzarasa2015Emergence}. Bursts denote a number of events occurring in short active periods separated by long inactive periods, which can be related to $1/f$ noise~\cite{Bak1987Selforganized, Weissman19881f, Ward20071f}. In general, non-Poissonian temporal patterns can be fully understood both by inhomogeneous interevent times (IETs) and by correlations between IETs~\cite{Jo2017Modeling}. Inhomogeneous and correlated IETs have been called \emph{correlated bursts}~\cite{Karsai2012Universal, Jo2018Limits}. Then, along with the information on who interacts with whom, one can comprehensively characterize the topological and temporal interaction structure of complex systems~\cite{Gauvin2018Randomized}.

Non-Poissonian bursty interactions between constituents of a system have been known to strongly affect the dynamical processes taking place in the system~\cite{Vazquez2007Impact, Iribarren2009Impact, Karsai2011Small, Miritello2011Dynamical, Rocha2011Simulated, Rocha2013Bursts, Takaguchi2013Bursty, Masuda2013Predicting, Perotti2014Temporal, Jo2014Analytically, PastorSatorras2015Epidemic, Delvenne2015Diffusion}; in particular, spreading dynamics in temporal networks has been extensively studied. An important question is what features of temporal networks are most relevant to predict the speed of propagation, e.g., of disease or information. One of the crucial and widely studied features is the inhomogeneity of IETs in the temporal interaction patterns. It was shown that the bursty interaction patterns can slow down the early-stage spreading by comparing the simulated spreading behaviors in some empirical networks and in their randomized versions~\cite{Vazquez2007Impact, Karsai2011Small, Perotti2014Temporal}. The opposite tendency was also reported using another empirical network or model networks~\cite{Rocha2011Simulated, Rocha2013Bursts, Jo2014Analytically}. 

In contrast to the IET distributions, yet little is known about the effects of correlations between IETs on the spreading, except for few recent works~\cite{Artime2017Dynamics, Masuda2018Gillespie}. This could be partly because the contagion dynamics studied in many previous works, e.g., susceptible-infected (SI) dynamics~\cite{PastorSatorras2015Epidemic}, has focused on an immediate infection upon the first contact between susceptible and infected nodes, hence without the need to consider correlated IETs. In another work~\cite{Gueuning2015Imperfect}, probabilistic contagion dynamics, i.e., naturally involving multiple consecutive IETs, was studied by assuming inhomogeneous but uncorrelated IETs. Therefore, the effects of inhomogeneous and correlated IETs on the spreading need to be systematically studied for better understanding the dynamical processes in complex systems.

In our paper, we study the effects of inhomogeneous and correlated IETs on the spreading taking place in temporal networks, by incorporating two contagion models, i.e., two-step deterministic SI and probabilistic SI dynamics. For modeling the inhomogeneous IETs, we consider power-law distributions of IETs, denoted by $\tau$, as
\begin{equation}
    P(\tau)\sim \tau^{-\alpha},
\end{equation}
with power-law exponent $\alpha$~\cite{Karsai2018Bursty}. For characterizing the correlations between IETs, we adopt a memory coefficient $M$~\cite{Goh2008Burstiness} among others, e.g., bursty train sizes~\cite{Karsai2012Universal}. The memory coefficient $M$ for a sequence of $n$ IETs has been estimated by
\begin{equation}
    M \equiv \frac{1}{n - 1}\sum_{i=1}^{n-1}\frac{(\tau_i - \mu_1)(\tau_{i+1} - \mu_2)}{\sigma_1 \sigma_2},
    \label{eq:M_origin}
\end{equation}
where $\mu_1$ ($\mu_2$) and $\sigma_1$ ($\sigma_2$) are the average and the standard deviation of the first (last) $n-1$ IETs, respectively. Positive $M$ implies the tendency of large (small) IETs being followed by large (small) IETs. Negative $M$ points towards the opposite, while $M=0$ for uncorrelated IETs. We focus on the case with positive $M$ as evidenced by several empirical findings~\cite{Goh2008Burstiness, Wang2015Temporal, Guo2017Bounds, Bottcher2017Temporal}. In our setup, both $P(\tau)$ and $M$ are inputs of the model, requiring us to consider $M$ as a parameter rather than an estimator. Then by controlling the shape of $P(\tau)$ and the value of $M$ for interaction patterns between nodes, one can study the effects of correlated bursts on the spreading in temporal networks. By analyzing the contagion dynamics on a single link, and then by simulating the spreading in regular and random temporal networks, we conclude that the positively-correlated inhomogeneous IETs slow down the spreading.

Our paper is organized as follows: In Sec.~\ref{sec:models}, we describe our contagion models. In Sec.~\ref{sec:2DSI}, we study the two-step deterministic contagion dynamics by analyzing the case for a single-link setup and then by simulating the spreading in regular networks of infinite size. The same framework is applied to the probabilistic contagion dynamics in Sec.~\ref{sec:PSI}. After numerically studying both contagion dynamics in finite random networks in Sec.~\ref{sec:finite}, we conclude our paper in Sec.~\ref{sec:conclusion}.

\section{Models}\label{sec:models} 

\begin{figure}[tb]
    \centering
    \includegraphics[width=0.95\columnwidth]{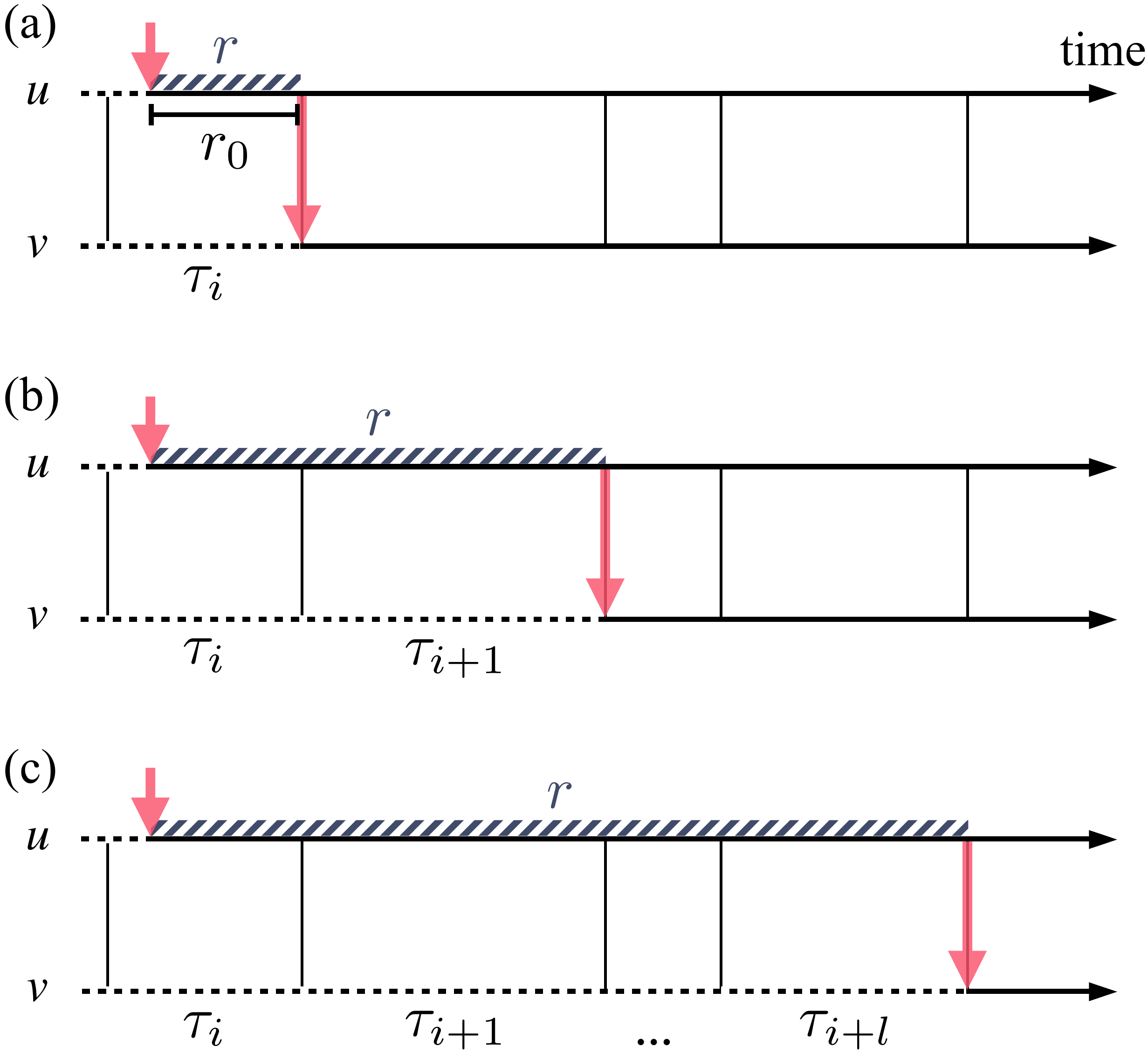}
    \caption{Schematic diagrams for (a) the one-step deterministic susceptible-infected (SI) dynamics, (b) the two-step deterministic SI dynamics, and (c) the probabilistic SI dynamics. For each node, the susceptible or intermediate state is represented by a dashed horizontal line, while the infected state is by a solid horizontal line. In each panel, a node $u$ gets infected in a random time, denoted by an upper vertical arrow, then it tries to infect its susceptible neighbor $v$ whenever they make contact (vertical lines). The successful infection of $v$ by $u$ is marked by a lower vertical arrow. The time interval between the infection of $u$ and that of $v$ (striped band) defines the transmission time $r$. For the definitions of $r_0$ and $\tau$s, see the text.}
    \label{fig:schematic}
\end{figure}

In order to study the spreading dynamics, we consider one of the extensively studied epidemic processes, i.e., susceptible-infected (SI) dynamics~\cite{PastorSatorras2015Epidemic}: A state of each node in a network is either susceptible ($S$) or infected ($I$), and an infected node can infect a susceptible node by the contact with it. Here we assume that the contact is instantaneous. The simplest SI dynamics could be one-step deterministic SI (1DSI) dynamics, where a susceptible node is immediately infected after its first contact with an infected node, see Fig.~\ref{fig:schematic}(a). This dynamics can be described by
\begin{equation*}
    S + I \to 2I.
\end{equation*}
In order to study the effect of correlations between IETs on the spreading in the simplest setup, we introduce two-step deterministic SI (2DSI) dynamics as a variation of generalized epidemic processes~\cite{Janssen2004Generalized, Dodds2004Universal, Bizhani2012Discontinuous, Chung2014Generalized}, see Fig.~\ref{fig:schematic}(b). Here a susceptible node first changes its state to an intermediate state ($S'$) upon its first contact with an infected node; it then becomes infected after the second contact with the same or another infected node. This can be written as
\begin{gather*}
    S + I \to S' + I, \\
    S' + I \to 2I.
\end{gather*}
To be more realistic, one can study a probabilistic SI (PSI) dynamics, in which the infection occurs with probability $\eta$ ($0 < \eta \leq 1$), as depicted in Fig.~\ref{fig:schematic}(c), i.e.,
\begin{equation*}
    S + I \xrightarrow{\eta} 2I.
\end{equation*}
The case with $\eta=1$ corresponds to the deterministic infection. In general, due to the stochastic nature of infection, the correlations between IETs in the contact patterns can influence the spreading behavior.

For modeling the interaction structure in a population, we focus on Bethe lattices as networks of infinite size, while regular random graphs and Erd\"{o}s-R\'{e}nyi random graphs will be later considered for finite network models. As for the temporal contact patterns, we assume that the contacts between a pair of nodes or on a link connecting these nodes are instantaneous and undirected. Moreover, the contact pattern on each link is assumed to be independent of the states of two end nodes as well as of contact patterns on other links. The contact pattern on each link is modeled by a statistically identical event sequence with inhomogeneous and correlated IETs. For this, the shape of IET distribution $P(\tau)$ and the value of memory coefficient $M$ are inputs of our model. Firstly, we consider a power-law IET distribution with a lower bound $\tau_{\min}$ and an exponential cutoff $\tau_\mathrm{c}$ as follows:
\begin{equation}
    P(\tau) = 
    \begin{cases}
    \frac{\tau_\mathrm{c}^{\alpha - 1}}{\Gamma(1 - \alpha, \tau_{\min} / \tau_\mathrm{c})} \tau^{-\alpha} e^{-\tau / \tau_\mathrm{c}} & \textrm{for}\ \tau\geq\tau_{\min},\\
    0 & \textrm{otherwise},
\end{cases}
    \label{eq:iet_distr}
\end{equation}
where $\Gamma$ is the upper incomplete Gamma function. We fix $\tau_{\min} = 1$ without loss of generality and set $\tau_\mathrm{c} = 10^3$ in our work, leaving us with one relevant parameter, i.e., the power-law exponent $\alpha$. Based on the empirical findings for $\alpha$~\cite{Karsai2018Bursty}, we consider the case with $1.5\leq \alpha\leq 3$. Secondly, only the positive memory coefficient $M$ is considered, precisely, $0\leq M< 0.4$, based on the empirical observations~\cite{Goh2008Burstiness, Wang2015Temporal, Guo2017Bounds, Bottcher2017Temporal}. 

In order to implement the inhomogeneous and correlated IETs for each link, we draw $n$ random values from $P(\tau)$ to make an IET sequence $T\equiv \{\tau_i\}_{i=1,\cdots,n}$, for sufficiently large $n$. Using the definition of Eq.~\eqref{eq:M_origin}, we measure the memory coefficient from $T$, denoted by $\tilde M$. Two IETs are randomly chosen in $T$ and swapped only when this swapping makes $\tilde M$ closer to $M$, i.e., the target value. By repeating the swapping, we can obtain the IET sequence whose $\tilde M$ is close enough to $M$. Finally the event sequence is obtained from the IET sequence.

\section{Two-step deterministic contagion}\label{sec:2DSI}

\subsection{Single-link transmission}\label{subsec:2Dsingle}

For understanding the spreading behavior on temporal networks, we first focus on how long it takes for the infection to transmit across a single link, say, from a node $u$ to its neighbor $v$. If $u$ gets infected from its neighbor than $v$ in time $t_u$, and later it infects $v$ in time $t_v$, the time interval between $t_u$ and $t_v$ defines the transmission time $r\equiv t_v-t_u$. Here we assume that $v$ is not affected by any other neighbors than $u$, for the sake of simplicity. In order for the infected $u$ to infect the susceptible $v$, $u$ must wait at least for the next contact with $v$. This waiting or residual time is denoted by $r_0$, see Fig.~\ref{fig:schematic}. For the one-step deterministic SI dynamics, $r=r_0$. Due to the independence of contact patterns of neighboring links, we can consider the infection of $u$ to occur at random in time, leading to the transmission time distribution as
\begin{equation}
    Q_{\rm 1D}(r) = \frac{1}{\mu} \int_{r}^\infty d\tau P(\tau),
    \label{eq:Qr_1D}
\end{equation}
with a finite $\mu$ denoting the mean IET. The average transmission time is directly obtained as
\begin{equation}
    \label{eq:avg_r_1D}
    \langle r\rangle_{\rm 1D} \equiv \int_0^\infty dr rQ_{\rm 1D}(r)= \frac{1}{2}\left(\mu + \frac{\sigma^2}{\mu}\right),
\end{equation}
where $\sigma^2$ denotes the variance of IETs~\cite{Karsai2018Bursty}. Note that a larger variance of IETs results in a larger average transmission time, expected to slow down the spreading.

In general, $r$ can be given as the sum of $r_0$ and subsequent IETs, as depicted in Fig.~\ref{fig:schematic}(b,~c), for the generalized epidemic processes, including our two-step deterministic SI (2DSI) dynamics. In the case with 2DSI dynamics, the transmission process involves two consecutive IETs. If the infection of $u$ occurs during the IET of $\tau_i$, then the transmission time is written as 
\begin{equation}
    \label{eq:r_2D}
    r=r_0+\tau_{i+1},
\end{equation}
with $\tau_{i+1}$ denoting the IET following $\tau_i$. Then the transmission time distribution is obtained as
\begin{equation}
    Q_{\rm 2D}(r) = \frac{1}{\mu} \int_0^r d\tau_{i+1} \int_{r-\tau_{i+1}}^\infty d\tau_i P(\tau_i,\tau_{i+1}),
    \label{eq:Qr_2D}
\end{equation}
where it is obvious from Eq.~\eqref{eq:r_2D} that $\tau_i\geq r_0=r-\tau_{i+1}$ and $\tau_{i+1}\leq r$. Information on the correlations between IETs is carried by the joint distribution function $P(\tau_i,\tau_{i+1})$. The average transmission time is calculated as 
\begin{equation}
    \langle r\rangle_{\rm 2D} \equiv \int_0^\infty dr rQ_{\rm 2D}(r)= \langle r\rangle_{\rm 1D}+\frac{1}{\mu} \langle \tau_i\tau_{i+1}\rangle,
\end{equation}
where
\begin{equation}
    \langle \tau_i\tau_{i+1}\rangle\equiv \int_0^\infty d\tau_i \int_0^\infty d\tau_{i+1} \tau_i\tau_{i+1} P(\tau_i,\tau_{i+1}).
    \label{eq:tautau1}
\end{equation}
In order to relate this result to the memory coefficient in Eq.~\eqref{eq:M_origin}, we define a parameter as
\begin{equation}
    M \equiv \frac{\langle \tau_i \tau_{i+1} \rangle - \mu^2}{\sigma^2}
    \label{eq:M_approx}
\end{equation}
to finally obtain the analytical result of the average transmission time: 
\begin{equation}
    \label{eq:avg_r_2D}
    \langle r\rangle_{\rm 2D} = \frac{3}{2} \mu + \left(\frac{1}{2} + M\right) \frac{\sigma^2}{\mu}.
\end{equation}
In the case with $M=0$ for uncorrelated IETs, one gets $\langle r\rangle_{\rm 2D} = \langle r\rangle_{\rm 1D}+\mu$.

We remark that our result in Eq.~\eqref{eq:avg_r_2D} is valid for arbitrary functional forms of IET distributions as long as their mean and variance are finite. $M$ is coupled with $\sigma^2/\mu$, implying that the impact of correlations between IETs becomes larger with broader IET distributions. More importantly, we find that a stronger correlation between consecutive IETs leads to a larger average transmission time. This can be understood in terms of the role of the variance of IETs in the average transmission time, as shown in the 1DSI case. That is, the variance of the sum of two consecutive IETs is amplified by the positive correlation between those IETs. Based on the result of the single-link analysis, the positive correlation between IETs is expected to slow down the spreading in a population. 

\subsection{Spreading in Bethe lattices}\label{subsec:2DBethe}

\begin{figure}[t!]
    \centering
    \includegraphics[width=\columnwidth]{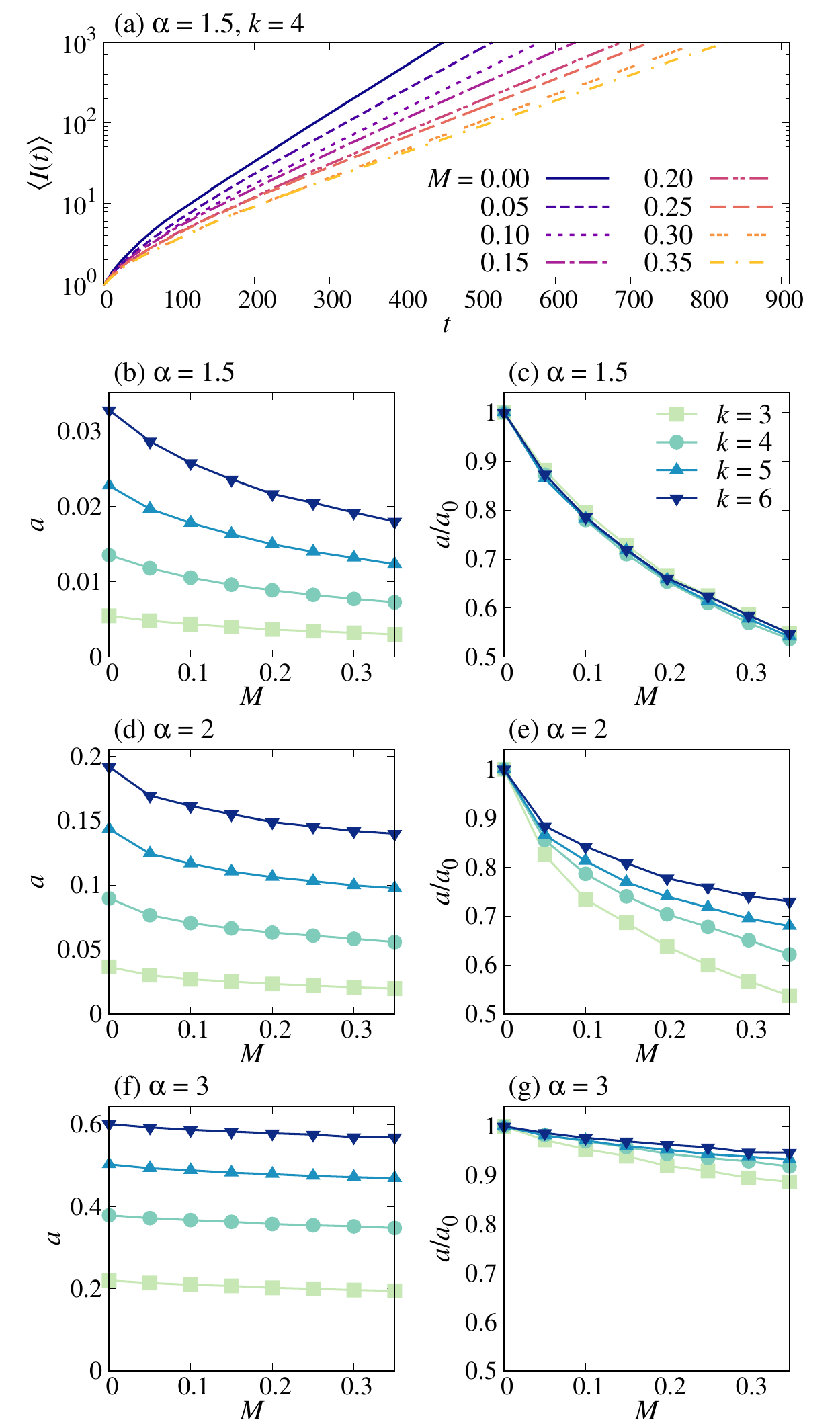}
    \caption{Two-step deterministic SI dynamics in Bethe lattices: (a) Average numbers of infected nodes as a function of time, $\langle I(t)\rangle$, in Bethe lattices with $k=4$ for the same IET distribution with power-law exponent $\alpha = 1.5$ in Eq.~\eqref{eq:iet_distr}, but with various values of memory coefficient $M$. Each point was averaged over $10^3$ runs with different initial conditions. (b--g) Estimated exponential growth rates $a$, defined in Eq.~\eqref{eq:spreading} (left panels) and their relative growth rates $a/a_0$ with $a_0\equiv a(M=0)$ (right panels) are plotted for various values of $k$, $\alpha$, and $M$. The lines are guides to the eye.}
    \label{fig:2dsi}
\end{figure}

In order to investigate the effects of correlations between IETs on the spreading in a population, we study spreading dynamics in a Bethe lattice, i.e., a regular tree of infinite size in which every node has exactly $k$ neighbors. One can relate this dynamics to the early-stage dynamics in regular random graphs, in which cycles are rare if the network size is sufficiently large. As mentioned, the contact pattern on each link is modeled by an independent and identical point process with the same $P(\tau)$ and $M$. Beginning with only one infected node in time $t=0$, we observe the number of infected nodes as a function of time. The average number of infected nodes $\langle I(t) \rangle$ is found to exponentially increase with time, e.g., as shown in Fig.~\ref{fig:2dsi}(a):
\begin{equation}
    \langle I(t) \rangle \sim e^{at}, 
    \label{eq:spreading}
\end{equation}
where $a=a(k,\alpha,M)$ denotes the exponential growth rate, known as the Malthusian parameter~\cite{Kimmel2002Branching}. $a$ turns out to be a decreasing function of $M$, indicating the slowdown of spreading due to the positive correlation between IETs, see Fig.~\ref{fig:2dsi}(b,~d,~f). The slowdown can be more clearly presented in terms of the relative growth rate $a/a_0$ with $a_0 \equiv a(M=0)$ for all cases of $k$ and $\alpha$, as shown in Fig.~\ref{fig:2dsi}(c,~e,~g). We summarize the main observations from the numerical simulations as follows:
\begin{enumerate}[(i)]
    \item $a$ decreases with $M$.
    \item $a$ increases with $\alpha$.
    \item $a$ increases with $k$.
    \item The deviation of $a/a_0$ from $1$ tends to be larger for smaller $\alpha$. 
\end{enumerate}
The observation (i) is expected from Eq.~\eqref{eq:avg_r_2D}, so is (ii) as both $\mu$ and $\sigma^2/\mu$ decrease with $\alpha$. (iii) is trivial. (iv) implies that the effect of $M$ becomes larger for smaller $\alpha$, which can be roughly understood by a larger value of $\sigma^2/\mu$ coupled to $M$ in Eq.~\eqref{eq:avg_r_2D}. We remark that Eq.~\eqref{eq:avg_r_2D} is the result for a single-link transmission, requiring us to study the transmission time in networks.

\begin{figure*}[t!]
    \centering
    \includegraphics[width=1.6\columnwidth]{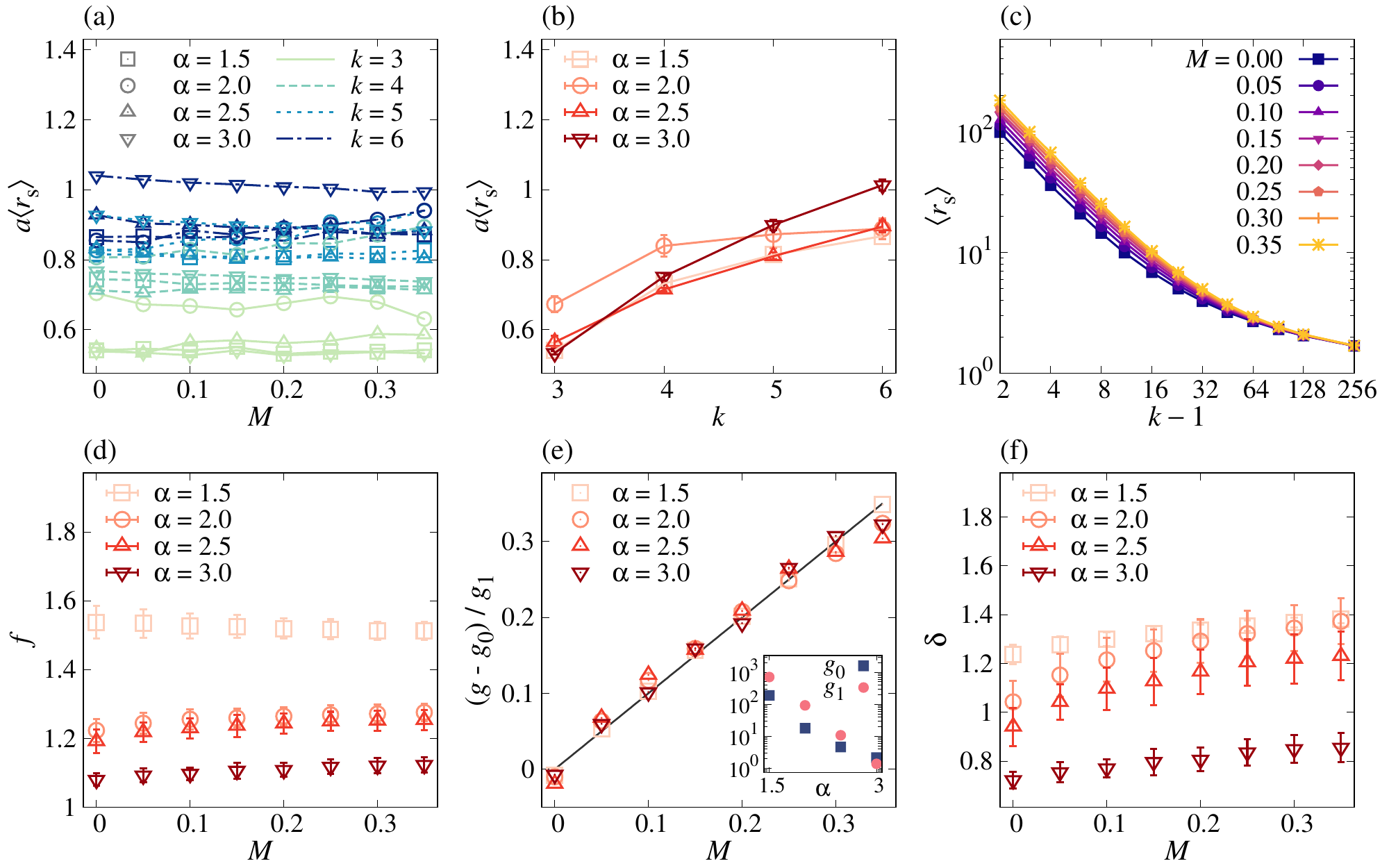}
    \caption{Two-step deterministic SI dynamics in Bethe lattices: The exponential growth rate $a$ can be explained in terms of the average shortest transmission time $\langle r_{\rm s}\rangle$, where $r_{\rm s}$ is defined in Eq.~\eqref{eq:shortest}. (a) $a\langle r_\mathrm{s} \rangle$ is overall independent of $M$ for all values of $k$ and $\alpha$. (b) $a \langle r_\mathrm{s} \rangle$, averaged over $M$, is an increasing function of $k$. The error bars represent standard deviation. (c) For the case of $\alpha=1.5$, $\langle r_{\rm s}\rangle$ is plotted as a function of $k-1$ for various values of $M$. Each point was averaged over $5\times10^{5}$ realizations. In (a--c), the lines are guides to the eye. (d--f) Using a functional form of $\langle r_{\rm s}\rangle = f+g(k-1)^{-\delta}$ in Eq.~\eqref{eq:shortest_hypo_general}, $f$, $g$, and $\delta$ are estimated for all values of $\alpha$ and $M$. In (e), using $g=g_0+g_1M$, we estimate $g_0$ and $g_1$ and plot $(g-g_0)/g_1$ against $M$, compared to the black line of $y=x$. The estimated $g_0$ and $g_1$ are shown in the inset of (e).}
    \label{fig:2dsi_anal}
\end{figure*}
	
In order to comprehensively understand the above observations, in particular, the $k$-dependence of $a$, we need to study the effect of time-ordering between infections to different neighbors~\cite{Lambiotte2013Burstiness}. For this, we introduce the shortest transmission time as
\begin{equation}
    r_{\rm s}\equiv \min \{r^{(1)},\cdots,r^{(k-1)}\},
    \label{eq:shortest}
\end{equation}
where $r^{(j)}$ for $j=1,\cdots,k-1$ denotes the transmission time from an infected node to its $j$th neighbor. Here we focus on the average of $r_{\rm s}$, denoted by $\langle r_{\rm s}\rangle$, which is a function of $k$, $\alpha$, and $M$. In Fig.~\ref{fig:2dsi_anal}(a), we numerically find that $a\langle r_{\rm s}\rangle$ is independent of $M$, implying that the effect of the correlations between IETs on spreading can be fully understood by $\langle r_{\rm s}\rangle$. Then we write $a$ as follows:
\begin{equation}
    a = \frac{h(k, \alpha)}{\langle r_\mathrm{s} \rangle}.
    \label{eq:malthus_hypo}
\end{equation}
Here $h(k,\alpha)$ is generally expected to be a function of $k$ and $\alpha$, while only its $k$-dependence is clearly shown in Fig.~\ref{fig:2dsi_anal}(b), where $h(k,\alpha)$ increases with $k$. In Fig.~\ref{fig:2dsi_anal}(c) we observe that as $k$ increases, $\langle r_{\rm s}\rangle$ algebraically decays before converging to a constant, enabling us to assume that
\begin{equation}
    \langle r_\mathrm{s} \rangle = f + g (k - 1)^{-\delta},
    \label{eq:shortest_hypo_general}
\end{equation}
where $f$, $g$, and $\delta$ are non-negative constants independent of $k$. By fitting the numerical results of $\langle r_{\rm s}\rangle$ using Eq.~\eqref{eq:shortest_hypo_general}, we find how these constants depend on $\alpha$ and $M$, as summarized in Fig.~\ref{fig:2dsi_anal}(d--f).

Firstly, we find that $f$ is overall independent of $M$. In the limit of $k \to \infty$, $\langle r_\mathrm{s} \rangle$ should asymptotically approach the smallest possible transmission time, denoted by $r_{\min}$, leading to $f=r_{\min}$. For the 2DSI dynamics and by our setup, $f=\tau_{\min}=1$ is expected, while the estimated values of $f$ show systematic deviations from $1$, possibly due to finite-size effects of $k$. Secondly and most importantly, $g$ turns out to linearly increase with $M$ such that
\begin{equation}
    g=g_0+g_1M,
\end{equation}
with positive coefficients $g_0$ and $g_1$, eventually leading to the linear dependence of $\langle r_{\rm s}\rangle$ in Eq.~\eqref{eq:shortest_hypo_general} on $M$. Moreover, both $g_0$ and $g_1$ are found to decrease with $\alpha$, shown in the inset of Fig.~\ref{fig:2dsi_anal}(e). These findings are comparable to the analytical result of average transmission time in Eq.~\eqref{eq:avg_r_2D}. Finally, the estimated values of $\delta$ appear to slightly increase with $M$, while we consider $\delta$ to be constant of $M$ in our argument. In sum, we rewrite $\langle r_{\rm s}\rangle$ in Eq.~\eqref{eq:shortest_hypo_general} as 
\begin{equation}
    \langle r_{\rm s}\rangle = r_{\min} + (g_0 + g_1 M) (k - 1)^{-\delta}.
    \label{eq:shortest_hypo}
\end{equation}
Combining $a$ in Eq.~\eqref{eq:malthus_hypo} and $\langle r_{\rm s}\rangle$ in Eq.~\eqref{eq:shortest_hypo}, we obtain the relative growth rate $a/a_0$ as
\begin{equation}
    \frac{a}{a_0} = \frac{r_{\min} + g_0 (k - 1)^{-\delta}}{r_{\min} + (g_0 + g_1 M) (k - 1)^{-\delta}},
\end{equation}
by which the observation (iv) can be understood: In one limiting case when $r_{\min} \ll g_0 (k - 1)^{-\delta}$, the relative growth rate is approximated as 
\begin{equation}
    \frac{a}{a_0}\approx \frac{g_0}{g_0+g_1M},
\end{equation}
which is independent of $k$ but clearly showing the $M$-dependence. This can explain the numerical findings in the case with small $\alpha$ in Fig.~\ref{fig:2dsi}(c). In the other limiting case when $r_{\min} \gg (g_0+g_1M) (k - 1)^{-\delta}$, one gets
\begin{equation}
    \frac{a}{a_0}\approx 1-\frac{g_1M}{r_{\min}}(k-1)^{-\delta},
\end{equation}
i.e., $a/a_0$ linearly but slightly decreases with $M$, showing a good agreement with the numerical results for large $\alpha$ in Fig.~\ref{fig:2dsi}(g). 

Conclusively, it turns out that the analytical result for the single-link transmission can to some extent account for the spreading behavior in networks, while more refined approach needs to be taken for better understanding the effect of network structure on spreading, e.g., $k$-dependence of $a$ in the case of Bethe lattices.

\section{Probabilistic contagion}\label{sec:PSI}

\subsection{Single-link transmission}\label{subsec:Psingle}

In a more realistic scenario than the two-step deterministic contagion dynamics, the infection can be described by a stochastic process, i.e., probabilistic SI (PSI) dynamics: An infected node infects a susceptible node with probability $\eta$ upon contact. Similarly to the deterministic cases in Sec.~\ref{sec:2DSI}, we begin with the analysis for a single-link transmission. The transmission time for a successful infection after $l$ failed attempts for $l \geq 0$ is
\begin{equation}
    r = 
    \begin{cases}
        r_0 & \text{for } l=0,\\
        r_0 + \sum_{j=1}^{l} \tau_{i+j} & \text{for } l > 0.
    \end{cases}
\end{equation}
The distribution of $r$, denoted by $Q_{\rm P}(r)$, can be written as the weighted sum of transmission time distributions for multi-step deterministic dynamics, similarly done in Ref.~\cite{Gueuning2015Imperfect}:
\begin{equation}
    Q_\mathrm{P}(r) = \sum_{l=0}^\infty \eta(1- \eta)^l Q_l(r),
    \label{eq:prob_transmission}
\end{equation}
where $Q_l(r)$ denotes the distribution of transmission time after $l$ failed attempts. Note that $Q_0(r)=Q_{\rm 1D}(r)$ in Eq.~\eqref{eq:Qr_1D} and $Q_1(r)=Q_{\rm 2D}(r)$ in Eq.~\eqref{eq:Qr_2D}. $Q_l(r)$ for general $l\geq 1$ is obtained as 
\begin{eqnarray}
    & Q_l(r) =& \frac{1}{\mu} \int_0^r dr_0 \prod_{j=1}^l \int_0^\infty d\tau_{i+j} \int_{r_0}^\infty d\tau_i P(\tau_i,\cdots,\tau_{i+l})\nonumber\\
    &&\times \delta \left(r-r_0-\sum_{j=1}^l \tau_{i+j}\right),
    \label{eq:prob_transmission_l}
\end{eqnarray}
where $\delta$ is a Dirac delta function, and $P(\tau_i,\cdots,\tau_{i+l})$ is the joint distribution function of $l+1$ consecutive IETs. Then one gets the average transmission time as follows:
\begin{equation}
    \langle r_l\rangle \equiv \int_0^\infty dr rQ_l(r)= \langle r\rangle_{\rm 1D}+\frac{1}{\mu} \sum_{j=1}^l \langle \tau_i\tau_{i+j}\rangle,
    \label{eq:rl}
\end{equation}
where
\begin{equation}
    \langle \tau_i\tau_{i+j}\rangle\equiv \prod_{j'=i}^{i+j} \int_0^\infty d\tau_{j'} \tau_i\tau_{i+j} P(\tau_i,\cdots,\tau_{i+j}).
    \label{eq:tautauj}
\end{equation}
For the details of the derivation, see Appendix~\ref{append:derive_rl}. We define the generalized memory coefficient between two IETs separated by $j-1$ IETs~\cite{Goh2008Burstiness} as
\begin{equation}
    M_j \equiv \frac{\langle \tau_i \tau_{i+j} \rangle - \mu^2}{\sigma^2},
    \label{eq:Mj}
\end{equation}
leading to
\begin{equation}
    \langle r_l\rangle = \langle r\rangle_{\rm 1D}+ l\mu + \frac{\sigma^2}{\mu} \sum_{j=1}^l M_j.
\end{equation}
\begin{figure}[t!]
    \centering
    \includegraphics[width=\columnwidth]{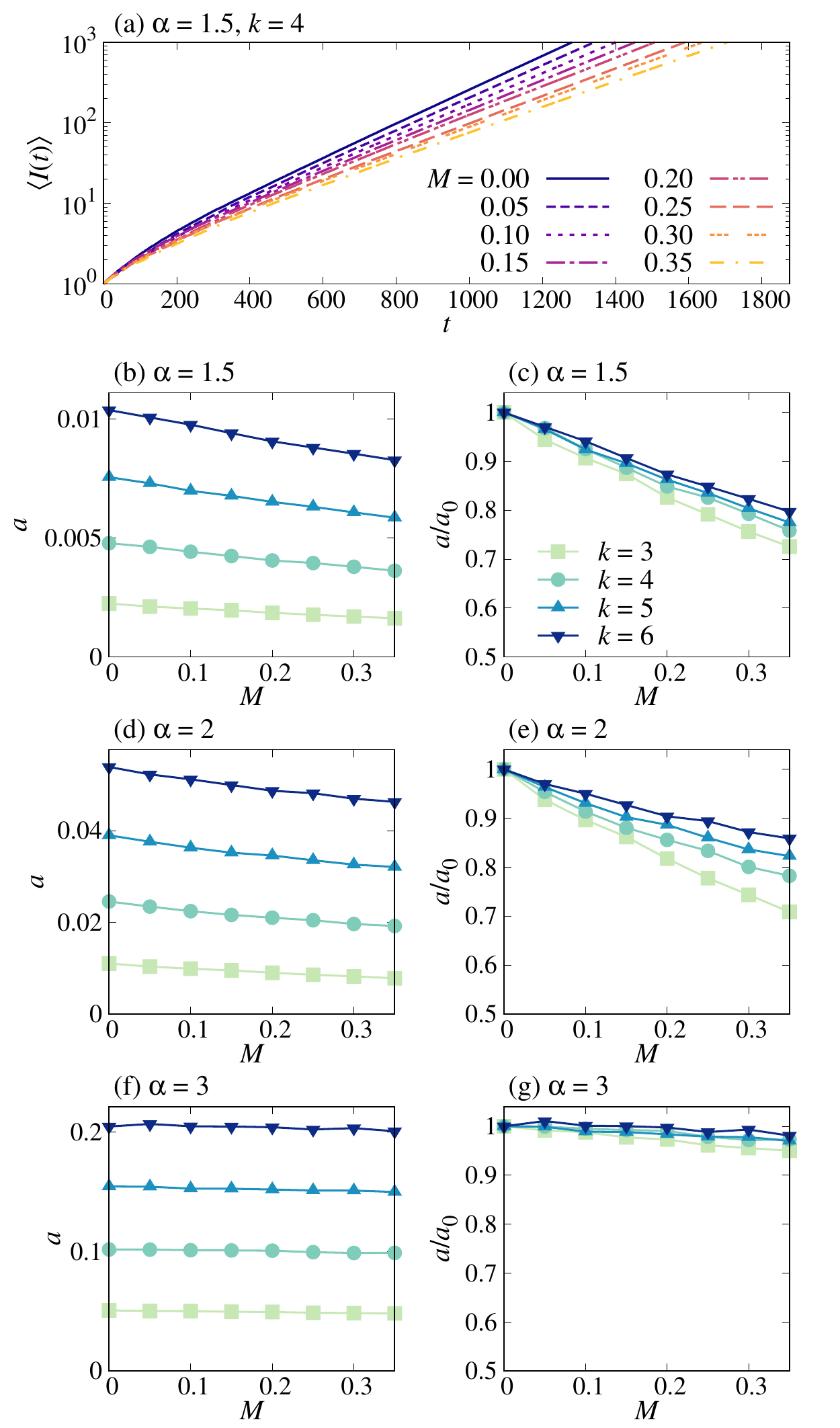}
    \caption{Probabilistic SI dynamics with $\eta=0.1$ in Bethe lattices. All notations and simulation details are the same as in Fig.~\ref{fig:2dsi}.}
    \label{fig:psi}
\end{figure} 
\begin{figure*}[t!]
    \centering
    \includegraphics[width=1.6\columnwidth]{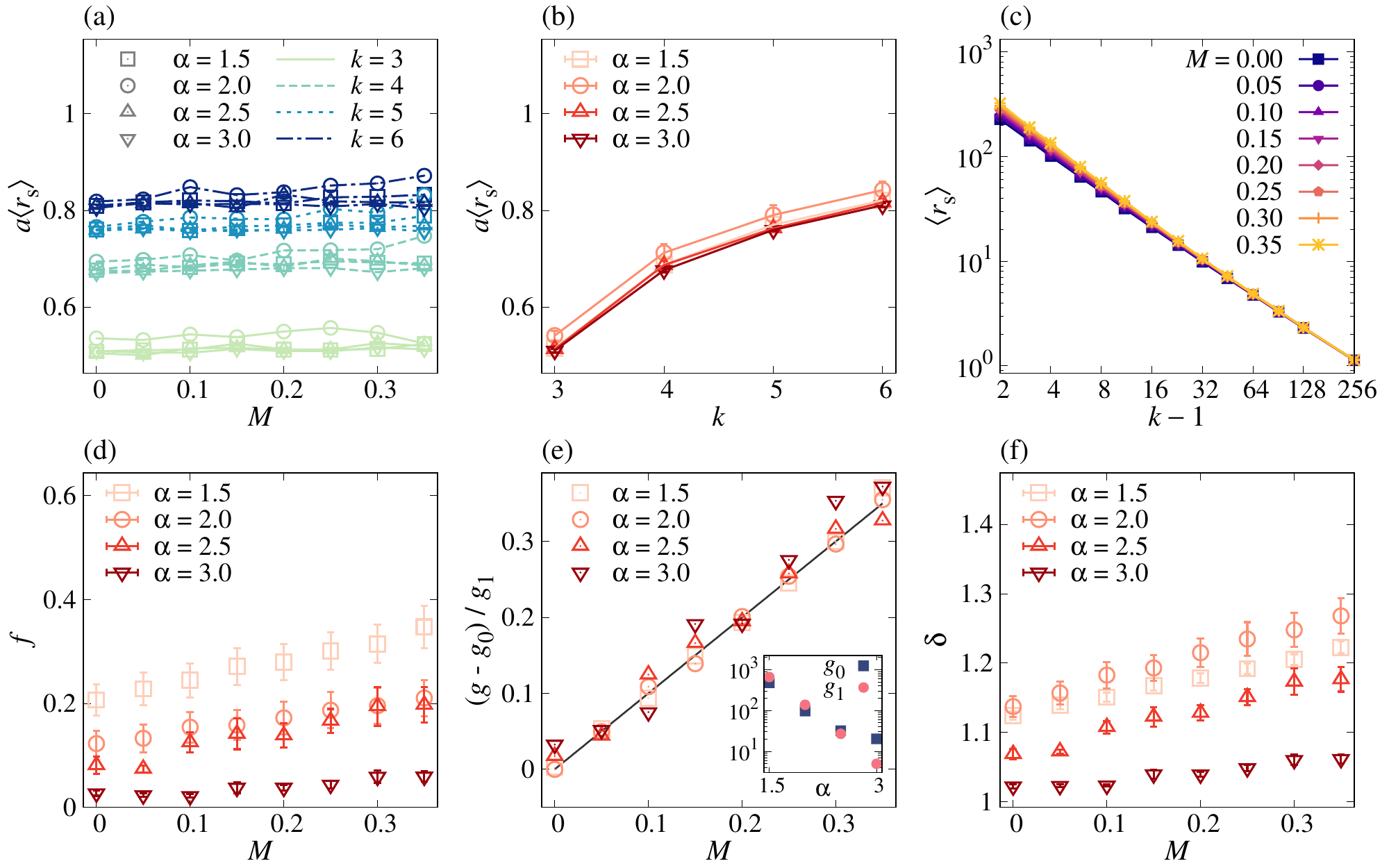}
    \caption{Probabilistic SI dynamics with $\eta=0.1$ in Bethe lattices. All notations and simulation details are the same as in Fig.~\ref{fig:2dsi_anal}.}
    \label{fig:psi_anal}
\end{figure*}
We then obtain the analytical result of the average transmission time for the PSI dynamics as
\begin{eqnarray}
    & \langle r\rangle_{\rm P} &\equiv \int_0^\infty dr rQ_{\rm P}(r)
    = \sum_{l=0}^\infty \eta(1- \eta)^l \langle r_l\rangle \nonumber\\
    &&=\langle r\rangle_{\rm 1D} + \frac{1-\eta}{\eta}\mu +\frac{\sigma^2}{\mu} \sum_{l=0}^\infty \eta(1- \eta)^l \sum_{j=1}^l M_j.
\end{eqnarray}
As we introduce only the correlations between two consecutive IETs in our model, we expect $M_j$ to exponentially decay according to $j$, where the decaying coefficient is denoted by $\gamma$ with $|\gamma|<1$: $M_j = \gamma M_{j-1} = \dots = \gamma^{j-1} M_1$, where $M_1=M$ in Eq.~\eqref{eq:M_approx}. Finally, we have 
\begin{equation}
    \langle r \rangle_{\rm P} = \left(\frac{1}{2}+ \frac{1- \eta}{\eta} \right) \mu + \left[\frac{1}{2}+ \frac{M(1- \eta)}{1- \gamma(1 - \eta)} \right] \frac{\sigma^2}{\mu}.
    \label{eq:avg_r_P}
\end{equation}
We note that this result is valid for arbitrary functional forms of IET distributions as long as their mean and variance are finite. Similarly to the deterministic case in Eq.~\eqref{eq:avg_r_2D}, the average transmission time for the PSI case turns out to be a linearly increasing function of the memory coefficient $M$. 

\subsection{Spreading in Bethe lattices}\label{subsec:PBethe}

Next, we numerically examine the spreading behavior for the PSI dynamics with $\eta<1$ in Bethe lattices. Similarly to the two-step deterministic case, we observe an exponential growth in the average number of infected nodes as well as the slowdown of spreading when the memory coefficient is positive. For example, the case with $\eta=0.1$ is depicted in Fig.~\ref{fig:psi}. As $\eta$ approaches $1$, the slowdown effect due to the correlated IETs becomes weak, as expected (not shown). Based on the results in Fig.~\ref{fig:psi_anal}, we make overall the same conclusions as in the 2DSI case: $a$ is a decreasing (increasing) function of $M$ (both $\alpha$ and $k$), and the deviation of $a/a_0$ from $1$ tends to be larger for smaller $\alpha$. 

The above observations in the PSI case can be understood by the same argument made in Subsec.~\ref{subsec:2DBethe}, namely, the functional form of $a$ in Eq.~(\ref{eq:malthus_hypo}) with $\langle r_{\rm s}\rangle$ in Eq.~(\ref{eq:shortest_hypo}), but with some important differences: Firstly, the shortest possible transmission time is $r_{\min}=0$, although the estimated values of $f$ show systematic deviations from $0$ in Fig.~\ref{fig:psi_anal}(d). This deviation is denoted by a small positive value $\epsilon$, leading to $f=\epsilon$. Secondly, $\delta$ appears to be an increasing function of $M$ rather than a constant in Fig.~\ref{fig:psi_anal}(f), which we assume to be $\delta=\delta_0 +\delta_1M$ with positive coefficients $\delta_0$ and $\delta_1$. We therefore modify $\langle r_{\rm s}\rangle$ in Eq.~(\ref{eq:shortest_hypo}) as follows:
\begin{equation}
    \langle r_{\rm s}\rangle = \epsilon + (g_0 + g_1 M) (k - 1)^{-(\delta_0 + \delta_1 M)}.
    \label{eq:psi_shortest}
\end{equation}
We note that due to the positive $\delta_1$, the above $\langle r_{\rm s}\rangle$ may decrease with $M$ but only for sufficiently large $k$ and $M$. However, we find no evidence for the decreasing behavior in the ranges of $k$ and $M$ studied in our paper. Using Eq.~\eqref{eq:psi_shortest}, the relative growth rate is obtained as
\begin{equation}
    \frac{a}{a_0} = \frac{\epsilon + g_0 (k - 1)^{-\delta_0}}{\epsilon + (g_0 + g_1 M) (k - 1)^{-(\delta_0 + \delta_1 M)}}.
\end{equation}
Since $\epsilon$ is a small number, we consider only the case of $\epsilon \ll g_0 (k - 1)^{-\delta_0}$ to get the approximated relative growth rate as 
\begin{equation}
    \frac{a}{a_0}\approx \frac{g_0}{g_0+g_1M}(k-1)^{\delta_1 M},
\end{equation}
which can account for the $k$-dependence of $a/a_0$, including the case with $\alpha=1.5$ in Fig.~\ref{fig:psi}(c). In Fig.~\ref{fig:psi}(c,~e,~g), we observe that the difference between curves of $a/a_0$ for different $k$s increases and then decreases as $\alpha$ increases from $1.5$ to $3$. This non-monotonic behavior can be related to the non-monotonic behavior of $\delta_1$ as a function of $\alpha$, as depicted in Fig.~\ref{fig:psi_anal}(f). 

\section{Spreading in finite networks}\label{sec:finite}

\begin{figure*}[tb]
    \centering
    \includegraphics[width=\textwidth]{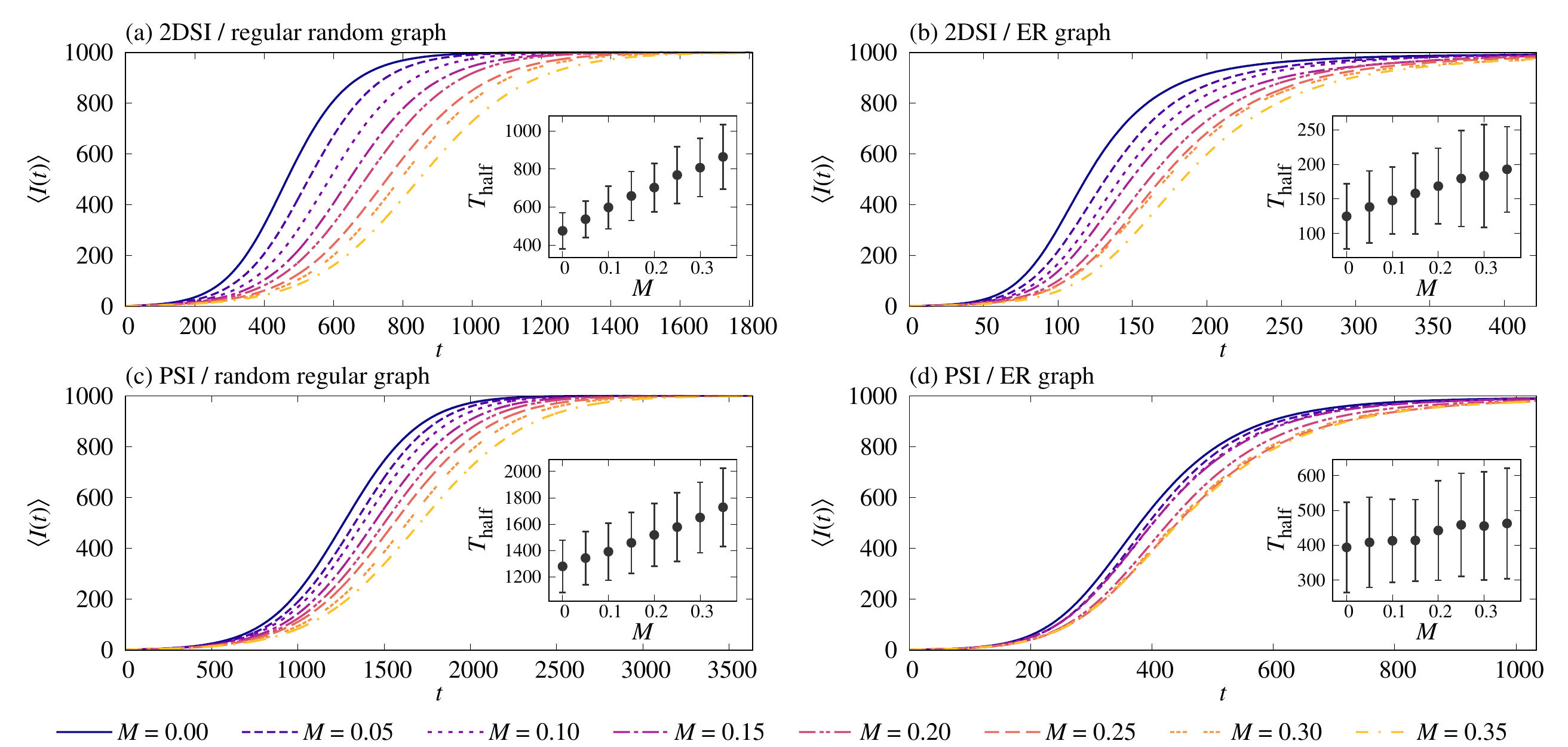}
    \caption{Average numbers of infected nodes as a function of time, $\langle I(t)\rangle$, for the two-step deterministic SI dynamics (a,~b) and the probabilistic SI dynamics with $\eta=0.1$ (c,~d) in two kinds of random networks of size $N = 10^3$: Regular random graphs with $k=4$ (a,~c) and Erd\"{o}s-R\'{e}nyi (ER) graphs with $p = 0.008$ (b,~d). In all cases, we used $\alpha=1.5$. Insets show the average and standard deviation of the time it takes to infect half of the population, $T_{\rm half}$.}
    \label{fig:finite}
\end{figure*}

So far we have focused on the spreading in Bethe lattices, i.e., regular networks of infinite size, which can also approximate the early-stage dynamics of spreading in finite networks as long as the cycles are rare. In addition to the early stage, the late-stage dynamics of spreading in finite networks has also been of interest~\cite{Min2011Spreading, Jo2014Analytically}. For this, we employ two network models of size $N$: Random regular graphs, in which every node has exactly $k$ neighbors, and Erd\"{o}s-R\'{e}nyi random graphs, in which every possible pair of nodes is connected with a probability $p$, hence the average degree is $p(N-1)$. On each of these graphs, both 2DSI and PSI dynamics are tested by the numerical simulations to measure the average numbers of infected nodes as a function of time, $\langle I(t)\rangle$. In all cases, we use networks of size $N=10^3$, and the results are averaged over $10^3$ simulation runs with different initial conditions for each parameter set.  

In Fig.~\ref{fig:finite} we find that the positive correlation between IETs lowers the average number of infected nodes for the entire range of time. This tendency can be quantified by the time it takes to infect half of the population, denoted by $T_{\rm half}$. The average value of $T_{\rm half}$ is increasing with $M$ for each parameter set, as depicted in the insets of Fig.~\ref{fig:finite}. This is consistent with the analytical results for the single-link transmission and with the numerical results for the spreading on Bethe lattices, leading to the conclusion that the positive correlation between IETs on each link slows down the spreading in a population. We note that in another work using the conditional distribution function $P(\tau_{i+1}|\tau_i)$~\cite{Artime2017Dynamics}, the positive correlation between IETs was reported to reduce the time it takes to reach the fully infected state. This finding is somehow in contrast to our conclusion, calling for more systematic approaches.

\section{Conclusions}\label{sec:conclusion}

Spreading dynamics in temporal networks has been extensively studied for tackling the important issue of what features of temporal networks are most relevant to the speed of spreading taking place in such networks. One of the widely studied features is the inhomogeneity of interevent times (IETs), typically represented by heavy-tailed IET distributions, in the temporal interaction patterns between nodes. Although the impact of the inhomogeneous IETs on the spreading has been largely explored, yet little is known about the effects of correlations between IETs on the spreading. It is partly because the contagion dynamics studied in many previous works focuses on the immediate infection upon the first contact between susceptible and infected nodes, hence without the need to consider the correlated IETs. However, since temporal correlations in the interaction patterns can be fully understood both by IET distributions and by correlations between IETs~\cite{Jo2017Modeling}, the effects of inhomogeneous and correlated IETs on the spreading need to be systematically studied for better understanding the dynamical processes in complex systems. For this, we consider two contagion dynamics, i.e., two-step deterministic SI and probabilistic SI dynamics, naturally involving multiple consecutive IETs. For both dynamics, we derive analytical expressions of average transmission times $\langle r\rangle$ for a single-link setup, which turn out to linearly increase with the memory coefficient $M$ as shown in Eqs.~\eqref{eq:avg_r_2D} and~\eqref{eq:avg_r_P}. Therefore, the positive correlation between IETs is expected to slow down the spreading. By performing numerical simulations of the contagion dynamics in regular networks of infinite size and random graphs of finite size, we conclude that the positive correlation between IETs indeed slows down the spreading, compared to the case of uncorrelated IETs but from the same IET distributions. 

The numerically obtained spreading speed, e.g., in Bethe lattices of degree $k$, could be successfully explained by means of the statistics of the shortest transmission time among $k-1$ transmission times from one infected node to its $k-1$ susceptible neighbors. In the case when IETs in the interaction patterns are largely homogeneous, the average transmission time $\langle r \rangle$ will serve as a representative timescale that determines the spreading speed. However, in the other case with inhomogeneous IETs or heavy-tailed IET distributions, the transmission time to each of $k-1$ neighbors will be heterogeneously distributed, implying that neighbors infected earlier tend to spread the disease or information more quickly, hence more broadly, than those infected later. In this sense, the majority of the infected nodes can be largely explained by the descendants of early-infected neighbors, and the characteristic timescale of spreading speed can also be set by the average shortest transmission time $\langle r_{\rm s}\rangle$, rather than $\langle r\rangle$. Unfortunately, as the analysis of $\langle r_{\rm s}\rangle$ appears not to be straightforward, more detailed and rigorous understanding of the behavior of $\langle r_{\rm s}\rangle$ is left for future works.

Finally, we remark that in addition to the memory coefficient, the correlations between IETs have also been identified by other methods, e.g., in terms of bursty trains, which can detect long-range memory effects between IETs~\cite{Karsai2012Universal}: The number of each bursty train, i.e., the burst size, has been described by heavy-tailed distributions. Regarding this, the relation between memory coefficient and burst size distributions was recently studied~\cite{Jo2018Limits}. Our approach can be extended by incorporating such heavy-tailed burst size distributions. We also note that more realistic network structures can be adopted for modeling temporal networks, such as networks with heterogeneous degrees~\cite{Barabasi1999Emergence} and community structure~\cite{Fortunato2010Community} among other network properties, e.g., stylized facts in social networks~\cite{Jo2018Stylized}.

\begin{acknowledgments}
The authors acknowledge financial support by Basic Science Research Program through the National Research Foundation of Korea (NRF) grant funded by the Ministry of Education (2015R1D1A1A01058958).
\end{acknowledgments}

\appendix

\section{Derivation of the average transmission time after $l$ failed attempts}\label{append:derive_rl}

The average of the transmission time after $l$ failed attempts, i.e., $\langle r_l\rangle$ in Eq.~\eqref{eq:rl} can be calculated using $Q_l(r)$ in Eq.~\eqref{eq:prob_transmission_l} as follows:
\begin{eqnarray}
    &\langle r_l\rangle =& \frac{1}{\mu} \int_0^\infty dr r \int_0^r dr_0 \prod_{j=1}^l \int_0^\infty d\tau_{i+j} \int_{r_0}^\infty d\tau_i \nonumber\\
    && P(\tau_i,\cdots,\tau_{i+l}) \delta \left(r-r_0-\sum_{j=1}^l \tau_{i+j}\right).
    \label{eq:rl_full}
\end{eqnarray}
We interchange the order of integration with respect to $r_0$ and $\tau_i$, i.e.,
\begin{equation}
    \int_0^r dr_0 \int_{r_0}^\infty d\tau_i = \int_0^\infty d\tau_i \int_0^{\min\{r,\tau_i\}} dr_0
\end{equation}
to rewrite Eq.~\eqref{eq:rl_full} as
\begin{eqnarray}
    &&\langle r_l \rangle = \frac{1}{\mu} \prod_{j=0}^l \int_0^\infty d\tau_{i+j} P(\tau_i,\cdots,\tau_{i+l})\times \nonumber\\
    && \int_0^\infty dr r \int_0^{\min\{r,\tau_i\}} dr_0 \delta \left(r-r_0-\sum_{j=1}^l \tau_{i+j}\right).
\end{eqnarray}
Calculation of the second line in the above equation is straightforward:
\begin{equation}
    \int_{\sum_{j=1}^l \tau_{i+j}}^{\sum_{j=0}^l \tau_{i+j}}dr r = \frac{\tau_i^2}{2}+\tau_i\sum_{j=1}^l \tau_{i+j},
\end{equation}
enabling us to finally obtain
\begin{equation}
    \langle r_l\rangle = \langle r\rangle_{\rm 1D} +\frac{1}{\mu} \sum_{j=1}^l \langle \tau_i\tau_{i+j}\rangle.
\end{equation}
Note that $\langle r\rangle_{\rm 1D} = \langle \tau^2\rangle / (2\mu)$ in Eq.~\eqref{eq:avg_r_1D}, with $\langle \tau^2\rangle$ denoting the second moment of the IET distribution $P(\tau)$.

\bibliographystyle{apsrev4-1}
%

\end{document}